\documentclass[11pt]{cernrep}
\usepackage{graphicx}
\usepackage{here}

\newcommand{\beeq}{\begin{equation}}
\newcommand{\eneq}{\end{equation}}
\newcommand{\beeqa}{\begin{eqnarray}}
\newcommand{\eneqa}{\end{eqnarray}}

\begin{document}
\title{
{\vspace{-1.2em} \parbox{\hsize}{\hbox to \hsize
{\hss \normalsize UPRF-2003-29, TRINLAT-03/07}}} \vglue 0.3 cm 
SUPERSYMMETRY ON THE LATTICE: WHERE DO PREDICTIONS AND RESULTS STAND ?}
\author{Alessandra~Feo \thanks{
To appear in the proceedings of Light-Cone Workshop: 
Hadrons and Beyond (LC03), Durham, England, 5-9 Aug 2003. 
This work was partially funded by the Enterprise-Ireland grant SC/2001/307. }}
\institute{School of Mathematics, Trinity College, Dublin 2, Ireland and \\
           Dipartimento di Fisica, Universit\`a di Parma and INFN Gruppo 
             Collegato di Parma, Parco Area delle Scienze, 7/A, 43100 Parma, Italy}

\maketitle
\begin{abstract}
I summarize recent results in lattice supersymmetry with special attention to $N=1$ Super Yang-Mills (SYM) 
theory. 
\end{abstract}

\section{INTRODUCTION}
Supersymmetric gauge theories exhibit fascinating phenomena which have attracted much 
attention since some results \cite{seiberg} about the ground state structure of some of 
these models have appeared.
For this reason, much more effort has been dedicated to formulating a lattice version 
of supersymmetric theories
(for some recent results, see for example, in the two dimensional Wess-Zumino model, 
\cite{catterall,campostrini,bietenholz}, 
four dimensional Wess-Zumino model \cite{fujikawa}, four dimensional $N=1$ SYM theory 
using Wilson fermions \cite{kirchner,campos,farchioni,montvay,feo2}, 
chiral fermions \cite{fleming} and other supersymmetric theories, as for example, 
in \cite{kaplan,kaplan2,giedt,itoh,so,hiller,pinsky}.
Unfortunately, I do not have time to cover all these subject and I refer the reader to 
\cite{feo}, for a complete list of references.

The dynamics of $N=1$ SYM is similar to QCD: confinement of the colored degrees of
freedom and spontaneous chiral symmetry breaking. 
The mechanism of spontaneous chiral symmetry breaking is quite peculiar in this model: 
$U(1)_\lambda $ is broken by the anomaly to $Z_{2 N_c}$ ($N_c$ is the number of colors) and it is 
spontaneously broken down to $Z_2$ due to a non-zero value of the gluino condensate \cite{witten}. 
Concerning the low-lying mass spectrum, there are various scenarios in the literature (with
different predictions) which can be directly investigated on the lattice in order to check their 
validity. The lattice is a fantastic laboratory which can help us to better understand 
whether supersymmetry is or not a symmetry of the Nature.
Also the investigation of the Ward-Takahashi identity (WTi) is a non-perturbative effect and has 
been studied on the lattice both, numerically and in lattice perturbation theory.
For the other way around, it is also interesting to study $N=1$ SYM far away from the 
supersymmetric limit in order to check the predictions of QCD for the glueball states, etc. 
Therefore, placing the QCD in a wider variety of theories as SYM can give us more information about it.

The outline of my talk is the following. First, an overview of the physics of the continuum $N=1$ SYM
theory together with some of the non-perturbative phenomena are presented. Then, the 
difficulties in formulating a lattice version of supersymmetric gauge theories are described.
Results and comparison when using either Wilson fermions or domain wall fermions 
are given.

\section{$N=1$ SYM CONTINUUM THEORY: NON-PERTURBATIVE RESULTS}
The continuum action density for $N=1$ SYM with a gauge group $SU(N_c)$ reads
\beeq
{\cal L} = -\frac{1}{4} F_{\mu \nu}^a(x) \, F_{\mu \nu}^a(x) + \frac{1}{2} \, \bar \lambda^a(x) \gamma_\mu 
({\cal D}_\mu \lambda(x))^a + m_{\tilde g} \bar \lambda^a(x) \lambda^a(x) \, ,
\label{action}
\eneq
where $\lambda^a$ is a 4-component Majorana spinor that satisfies the Majorana condition 
$\bar \lambda^a = {{\lambda}^{a}}^T C$. The gluon fields are represented by
$A_\mu = -i g A_\mu^a T^a $ and $ F_{\mu \nu} = -i g F_{\mu \nu}^{a} T^a$.
${\cal D}_\mu \lambda^a = \partial_\mu \lambda^a + g f_{abc} A_\mu^b \lambda^c$
is the covariant derivative in the adjoint representation. 

For $m_{\tilde g} = 0$ this action has a supersymmetry respect to the continuum 
supersymmetric transformations,
\beeqa
&& \hspace{-0.7 cm} \delta A_\mu(x) = -2 g \bar\lambda(x) \gamma_\mu \varepsilon \, ,  \nonumber \\
&& \hspace{-0.7 cm} \delta \lambda(x) = -\frac{i}{g}\sigma_{\rho\tau} F_{\rho\tau}(x) \varepsilon \, , \nonumber \\
&& \hspace{-0.7 cm} \delta \bar\lambda(x) = \frac{i}{g} \bar \varepsilon \sigma_{\rho\tau}F_{\rho\tau}(x) \, , 
\label{transform}
\eneqa
where $\sigma_{\rho\tau} = \frac{i}{2} [\gamma_\rho,\gamma_\tau] $, $\lambda = \lambda^a T^a$ 
and $\varepsilon $ is a global Grassmann parameter with Majorana properties.

These transformations relate fermions with bosons, they leave the action invariant, which means that
the variation of the lagrangian under these tranformations is proportional to a total derivative of a current
(then, the variation of the action is zero).
Also, they commute with the gauge transformations so that the resulting Noether current, $S_\mu$,
is gauge invariant. For $N=1$ SYM theory the supercurrent is given by,
\beeq
S_\mu(x) = - F_{\rho \tau}^a(x) \sigma_{\rho \tau} \gamma_\mu \lambda^a(x) \, .
\eneq

It is also expected that the supercurrent satisfy a renormalized supersymmetric WTi, 
$\partial_\mu S_\mu^R = 2 m_R \chi_R$, where $\chi_R = Z_\chi \chi$,  
and $\chi \equiv \frac{1}{2} F_{\rho \sigma}^a \sigma_{\rho \sigma} \lambda^a$ 
and $m_R$ is the renormalized gluino mass. 
We have supersymmetry for $m_R=0$, while the non-vanishing of $m_R$ describes a 
soft breaking of supersymmetry (this can be easily seen applying the supersymmetric transformations over the 
mass term in the density action: We see that it is not possible to write down this term as a total derivative of
an operator). It is generally assumed that supersymmetry is not anomalous and that only 
the mass term is responsible for a soft breaking. In \cite{shamir} however, the question of whether 
non-perturbative effects may raise a supersymmetry anomaly has been investigated. Only a study of 
the continuum limit of the lattice supersymmetric WTi can shed some light on this question.
I will be back to this point later on.

\subsection{Chiral symmetry breaking}
A non-zero gluino mass term in Eq.~(\ref{action}), 
${\cal L}_{mass} = m_{\tilde g}\bar \lambda^a \lambda^a $,
breaks supersymmetry softly (that means that all the non-renormalization theorem and cancellation 
of divergencies are preserved). 
In the massless case, the global chiral symmetry is $U(1)_{\lambda}$, which is broken by the anomaly,
\beeq
\partial_\mu J^5_\mu = \frac{N_c g^2}{32 \pi^2} \varepsilon^{\mu \nu \rho \sigma} F_{\mu \nu}^a 
F_{\rho \sigma}^a
\eneq
(where $J_\mu^5 = \bar \lambda \gamma_\mu \gamma_5 \lambda$), 
leaves a $Z_{2 N_c}$ subgroup unbroken. This $Z_{2 N_c}$ symmetry group is itself broken down to $Z_2$ due
to the non-zero value of the gluino condensate. 
The conseguence of this spontaneous chiral symmetry breaking is a 
first order phase transition at $m_{\tilde g} = 0$. That means the existence of $N_c$ degenerate ground
states with different orientations of the gluino condensate $(k=0,\cdots,N_c -1)$,
\beeq
\bigg< \lambda^a_\alpha \lambda^{a \alpha} \bigg> = c \Lambda^3 e^{\frac{2 \pi i k}{N_c}} \, .
\label{gluinocondensate}
\eneq
Here $\Lambda$ is the dynamical scale of the theory which can be calculated on the lattice,
for example, while $c$ is a numerical constant which depends on the renormalization scheme used to compute
$\Lambda$. Eq.~(\ref{gluinocondensate}) shows up the dependence on the gauge group.
For $SU(2)$, two degenerate ground states with opposite sign of the gluino condensate 
appear \cite{kirchner}, while for $SU(3)$,
we have three degenerate ground states at $k=k_c$ (for a first numerical study see \cite{su3}). 

\subsection{The value of the gluino condensate}
The value of the gluino condensate in the pure $N=1$ SYM theory has been calculated in the eighties 
by using two different methods. One is based on strong coupling instanton calculations,\cite{strong},
while the second one is based on weak coupling instanton calculations, \cite{weak}. They 
give different result for $c$ (in Eq.~(\ref{gluinocondensate})) and this was known as the $\frac{4}{5}$ puzzle.  
Various discussion about the validity of both methods can be found in the literature \cite{amati}.
More recently, a third elegant method \cite{davies} calculates the gluino condensate directly in the 
semiclassical approximation. This method gives results in agreement with the weak coupling instanton 
approximation \cite{weak,finnell} and confirm the correctness of the weak coupling instanton 
result. 

\subsection{Light hadron spectrum}
$N=1$ SYM theory describes interactions between gluons and gluinos. In analogy with QCD one expect 
that the spectrum of the model consists of colorless bound states of those fundamental excitations,
namely glueballs ($gg$), gluinoballs ($\lambda \lambda$) and gluino-glueball ($g \lambda$).
These fields can combined to form the chiral supermultiplet, 
\beeq
S(y) = \phi(y) + \sqrt{2} \theta \chi(y) + \theta^2 F(y) \, ,
\eneq
where $\phi$ represent the scalar and pseudoscalar gluinoballs while $\chi$ is their fermionic partner.
$F$ is an auxiliary field. It is tempting to think that $F$ represent the 
scalar and pseudoscalar glueballs when treating it as a dynamical field in the minimal 
Veneziano and Yankielowicz (VY) action, but this cause paradoxes \cite{sannino}.
For $N=1$ SYM theory a low energy effective action has been proposed by VY \cite{veneziano},
\beeq
{\cal L}_{eff}= \frac{1}{\alpha} (S^\dagger S)^{1/3} \vert_D + \gamma [(S \mbox{log} \frac{S}{\mu^3}
- S) \vert_F + h.c. ] \, .
\eneq
Expanding the effective VY action around its mininum, it is found that the low energy spectrum
forms a supermultiplet, consisting of a scalar meson $\bar \lambda^a \lambda^a$
(the adjoint$-f_0$), a pseudoscalar meson $\bar \lambda^a \gamma_5 \lambda^a$ (the adjoint$-\eta \prime$), 
and a spin-$1/2$ gluino-glueball particle, the $\chi$ . 
Glueballs are absent in this formulation. In the supersymmetric limit these masses are degenerate,
while the introduction of a $m_{\tilde{g}} \ne 0$ breaks supersymmetry softly and leads to
a splitting of the multiplet.
Generalization of the VY effective action in order to include glueballs are discussed, for example, in
\cite{farrar,cerdeno}. These generalizations give results which are in substantial agreement with 
lattice simulation using Wilson fermions \cite{campos,peetz}.

\section{LATTICE FORMULATION OF $N=1$ SYM THEORY} 
There are several important motivations for defining a field theory on the lattice. 
The most important one is to regularize a ``perturbative divergent theory" in a 
non-perturbative way. However, formulating a supersymmetric field theory 
on the lattice seems to be a difficult task. 
The first difficulty has to do with the failure of the Leibniz rule \cite{dondi}.
The lattice version of the Leibniz rule is order $O(a)$, thus, breaks supersymmetry.
In fact, if we discretize in a trivial way the Eqs.~(\ref{action},\ref{transform})
we see that the variation of the lagrangian under the 
discretized lattice transformations is not proportional to a total lattice derivative.
Only in the continuum limit
one can recover the continuum results plus terms order $O(a)$ (depending on the discretization 
prescription). 
This is not a severe problem due to the fact that terms that break 
supersymmetry are irrelevant, i.e., they cancel in the continuum limit and therefore there is 
no need for fine tuning in order to eliminate their contributions.

Another problem are the scalar mass terms in the supersymmetric theory that breaks supersymmetry. 
Since these operators are relevant, fine tuning is necessary in order to cancel their contributions.
In the case of $N=1$ SYM theory only one scalar mass term is present which has to be 
fine tuning. But in the case of extended supersymmetries, there will be a 
plethora of scalar fields and the situation would become drammatic if one would have to do 
a fine tuning for each of these operators.

The last problem is also well known in QCD. A naive regularization of fermions originates the 
doubling problem \cite{nielsen}.  This is also problematic for supersymmetry not only because 
the discretization results in a wrong number of fermions (16 instead of 1), but also 
results in a violation of the balance between bosons and fermions as supersymmetry requires. 
This problem can be treated as in QCD. In the following, I summarize some nice numerical results 
using the Wilson formulation and the domain wall approach (for a detailed review on these subjects, 
see \cite{feo,montvay,vranas2}).

\subsection{$N=1$ SYM theory with Wilson fermions}
In the Wilson formulation of Curci and Veneziano \cite{curci}, it is proposed to construct a non supersymmetric
discretized $N=1$ SYM theory with a supersymmetric continuum limit. 
In this formulation, supersymmetry is broken by the lattice itself, by the Wilson term and a soft breaking due 
to the gluino mass is present. Supersymmetry is recovered in the contimuum limit by tuning the bare parameters $g$ 
and the gluino mass $m_{\tilde{g}}$ (through the hopping parameter) to the supersymmetric limit.
The supersymmetric (and chiral) limit are both recovered simultaneously at $m_{\tilde{g}}=0$. 

The strategy using Wilson fermions is the following. By studying the pattern of chiral symmetry breaking, 
through the study of the first order phase transition of the gluino condensate it is then 
possible to determine the value of the critical 
hopping parameter which correspond to the supersymmetric limit ($m_{\tilde{g}}=0$). 
In \cite{kirchner,su3}, for a fixed value of 
$\beta$, it is introduced a gluino mass term that breaks supersymmetry and then it is tuned in order 
recovered supersymmetry in the continuum limit. 
At the supersymmetric (chiral) limit, a first order phase transition (or a crossover) should show up as a 
two double peak structure in the distribution of some order parameter (the gluino condensate, in this case), 
indicating that the corresponding 
$k_c$ is the critical hopping parameter corresponding to the supersymmetric limit. 
How do we know we are restoring supersymmetry in the continuum limit?
\begin{itemize}
\item
This can be achive for example, by investigating the low-lying mass spectrum and comparing with theoretical 
predictions. 
In \cite{campos,peetz}, the unquenched simulations using the multi-bosonic algorithm proposed by L\"uscher 
\cite{luscher} with 
a two-step variant (the TSMB) algorithm \cite{montvay2}, near the value of $k_c$, have been performed. 
An accurate study of this issue is non-trivial not only from the computational point of view but also due to 
several different theoretical formulations. An independent method for checking the supersymmetry restoration 
would be demanding. 
\item
Another independent way to study the supersymmetry restoration is through the study of the supersymmetric 
WTi.
This has been achieved both numerically \cite{farchioni} and in lattice perturbation theory \cite{feo2,taniguchi}.

\end{itemize}

\subsection{$N=1$ SYM theory with Domain Wall fermions}
$N=1$ SYM theory has been also studied on the lattice using the domain wall fermion (DWF) approach. 
Application of DWF in supersymmetric theories has been explored in \cite{neuberger}.
The DWF approach is defined by extending the space-time
to five dimensions. Also a non-zero five dimensional mass or domain wall height $m_0$, which 
controls the number of flavors, is present. 
The size of the fifth dimension, $L_s$, and free boundary conditions for the fermions are 
implemented. As a result, the two chiral components of the Dirac fermion are separated with 
one chirality bound exponentially on one wall and the other on the opposite wall.
For any value of $a$ the two chiralities mix only by an amount that decreases exponentially 
as $L_s \to \infty$. For $L_s = \infty$, the chiral symmetry is expected to be exact even 
at finite lattice spacing. Therefore, there is no need for fine tuning.
DWF offer the oportunity to separate the continuum limit, $a \to 0$, from the chiral limit, $L_s \to \infty$. 

First Monte Carlo simulations \cite{fleming} for $N=1$ $SU(2)$ SYM with DWF, using the HMDR \cite{gottlieb},
indicate the formation of a gluino condensate which is sustained at the chiral limit. 
The condensate is non-zero even for small volume and small supersymmetry breaking mass \cite{fleming}.
DWF have two welcomed properties that Wilson fermions do not have: There is no need for fine tuning.
The second one has to do with the sign of the Pfaffian which is positive while in the case of 
Wilson fermions this is not assured 
\footnote{In Wilson fermions there are no serious sign problems when $k<k_c$ 
\cite{campos,farchioni}}.

For the other hand, DWF introduce two extra parameters: $L_s$ and $m_0$. 
These two parameters together with the four dimensional mass, $m_f$, control the effective
fermion mass $m_{eff}$. In the free theory one find,
\beeq
m_{eff} = m_0 (2-m_0) [ m_f + (1 - m_0)^{L_s}] \, , 
\label{mass}
\eneq
with $0 < m_0 < 2$. The value of $m_0 = 1$ is optimal because finite $L_s$ effects do not contribute 
to $m_{eff}$. In the interactive theory, one would not expect such optimal value, due to the fact that $m_0$ will 
fluctuate. Then the goal would be to have $L_s$ large enough to have the second term of Eq.~(\ref{mass})
small, in order to simulate at reasonably small masses and extrapolate to the chiral limit,
$m_f \to 0$ and $L_s \to \infty$ \cite{vranas2}. There is a sort of ``fine tuning" in order to achive the better 
value in Eq.~(\ref{mass}). DWF are much more expensive than Wilson fermions from the computational
point of view. Wilson fermions are less expensive for SYM than for QCD.

\subsection{The supersymmetric Ward-Takahashi identity (WTi)}
The numerical simulations of the WTi \cite{farchioni} have been performed in order to determine
non-perturbatively a substracted gluino mass and the mixing coefficients of the supercurrent.
Thus, the supersymmetric WTi reads \cite{farchioni},
\beeq
\big< {\cal O}(y) \nabla_\mu S_\mu(x) \big> +
Z_T Z_S^{-1} \big< {\cal O}(y)\nabla_\mu T_\mu(x) \big> = m_R Z_S^{-1} \big< {\cal O}(y) \chi(x) \big> \, ,
\label{numerical}
\eneq
which can be computed at fixed $\beta$ and $k$. By choosing two elements of the $4 \times 4$ matrices,
having previously choosing the operator insertion ${\cal O}(y)$ in Eq.~(\ref{numerical}), 
a system of two equations can be solved for $Z_T Z_S^{-1}$ and $m_R Z_S^{-1}$ \cite{farchioni}.
The results show a restoration of supersymmetry in the continuum limit up to $O(a)$ effects. 
The vanishing gluino mass, extrapolated when determine $m_R Z_S^{-1}$, occurs at a value 
of the hopping parameter 
in agreement to the one calculated using the chiral symmetry breaking \cite{kirchner}. 

The supersymmetric WTi has been also studied in lattice perturbation theory \cite{feo2,taniguchi}.
The idea is to use an independent way to determine the value of $Z_T/Z_S$, which is equal to 
$Z_T/Z_S \equiv {Z_T}_{1-loop}$, in lattice perturbation theory 
\footnote{That is because, $Z_S = 1 + O(g_0^2)$ and $Z_T = O(g_0^2)$.}.
The starting point of the calculation is the renormalized supersymmetric WTi on the lattice,
introduced in Ref.~\cite{feo2},
\beeqa
&& Z_S \big< O \nabla_\mu S_\mu(x) \big> + Z_T \big< O \nabla_\mu T_\mu(x) \big> - 
2 (m_0 - \tilde{m}) Z_{\chi}^{-1} \big< O \chi^R(x) \big> + \nonumber \\
&& Z_{CT} \big< \frac{\delta O} {\delta \bar \xi(x)}|_{\xi = 0} \big> - 
Z_{GF} \big< O \, \frac{\delta S_{GF}}{\delta \bar \xi(x)}|_{\xi = 0} \big> -
Z_{FP} \big< O \, \frac{\delta S_{FP}}{\delta \bar \xi(x)}|_{\xi = 0} \big> + \sum_j Z_{B_j} \big< O B_j \big>=0 \, ,
\label{renorm}
\eneqa
where $\xi(x)$ is a localized transformation parameter, 
$ S_\mu(x) = -\frac{2 i}{g_0} \, \mbox{Tr} \, \big( G_{\rho \tau}(x) \sigma_{\rho \tau} \gamma_\mu \lambda(x) \big)$,
is a local definition of the lattice supercurrent which mixes with
$ T_\mu(x) = -\frac{2}{g} \, \mbox{Tr} \, \big( G_{\mu \nu}(x) \gamma_\nu \lambda(x) \big) $.
$\chi(x) = \frac{i}{g_0} \, \mbox{Tr} \, \big( G_{\rho \tau}(x) \sigma_{\rho \tau} \lambda(x) \big) $
is the gluino mass term, 
$\nabla_\mu $ is the symmetric lattice derivative,
$G_{\rho \tau}(x)$ is the clover plaquette operator and $\sigma_{\rho \tau} = \frac{i}{2} [\gamma_\rho,\gamma_\tau]$.
Then, $\big< \frac{\delta O} {\delta \bar \xi(x)}|_{\xi = 0} \big> $, 
$\big< O \, \frac{\delta S_{GF}}{\delta \bar \xi(x)}|_{\xi = 0} \big> $ and 
$\big< O \, \frac{\delta S_{FP}}{\delta \bar \xi(x)}|_{\xi = 0} \big> $ are the contact terms,
gauge fixing terms and Faddeev-Popov terms, respectively (we do not report them here, see Ref.~\cite{feo2}).
Notice that in Eq.~(\ref{renorm}) these terms are also renormalized. 
This is due to the fact that their one-loop corrections are not just multiples of the corresponding 
tree-level values.
$\sum_j Z_{B_j} \big< O B_j \big> $ represent the operator mixing, not only with 
non-gauge invariant operators (in the case the operator insertion $O$ is non-gauge invariant),
but also with gauge invariant operators which do not vanish in the off-shell regime
but vanish in the on-shell one. In principle, one would require a complete list of them, or, as in our case,
a sub-list of operators whose contributions are different from zero to the renormalization constant 
we are interested on ($Z_T$ in this case).

In the supersymmetric limit, the renormalized gluino mass is zero, so the third term of the first line 
in Eq.~(\ref{renorm}) vanish and we leave with a simple expression. From now on, when 
we refer to Eq.~(\ref{renorm}) we will assume this term equal to zero.
Considering now each matrix element in Eq.~(\ref{renorm}) with $O$ a non-gauge invariant operator
given by 
$O := A_\nu^b(y)\, \bar \lambda^a(z) $.
Each matrix element in Eq.~(\ref{renorm}) is proportional to each element of the $Gamma$-matrix base 
$\Gamma = \left\{ 1, \gamma_5, \gamma_\alpha, \gamma_5 \gamma_\alpha, \sigma_{\alpha \rho} \right\} $.
To determine $Z_T$ one needs the projections over $\gamma_\alpha$ and $\gamma_\alpha \gamma_5$.

In Fourier transformation (FT), we choose $p$ as the outcoming momentum for the gluon field 
$A_\mu$ and $q$ the incoming momentum for the fermion field $\lambda$. 
To calculate $Z_T$ one should pick up from each matrix element of Eq.~(\ref{renorm}) 
those terms which contains the same Lorentz structure 
as $S_\mu$ and $T_\mu$, to tree-level. Those operators which do not contain the same 
tree-level Lorentz structure as $S_\mu$ and $T_\mu$ do not enter in the determination of $Z_T$.

The renormalization constants as well as the operators, can be written as a power of $g_0$ \cite{feo2}, 
\beeqa
&& Z_{operator} = Z_{operator}^{(0)} + g_0^2 Z_{operator}^{(2)} + \cdots \, , \nonumber \\
&& \big< Operator \big> = \big< Operator \big>^{(0)} + g_0^2 \big< Operator \big>^{(2)} + \cdots \, ,
\label{operators}
\eneqa
where $\big< Operator \big>^{(2)}$, is the 1-loop correction
while $\big< Operator \big>^{(0)}$, is the tree-level value. 
To get $Z_T$ we need to distinguish the tree-level values of $S_\mu$ and $T_\mu$ and
for that reason we require general external momenta, $p$ and $q$.

Substituting Eq.~(\ref{operators}) into Eq.~(\ref{renorm}), up to order $g_0^2$, and
using the projections over $\gamma_\alpha$ and $\gamma_\alpha \gamma_5$ we obtain, respectively 
\cite{feo2}, 
\beeqa
&& \frac{1}{4} \mbox{tr} \big( \gamma_\alpha \big(\nabla_\mu S_\mu \big)^{(2)}_{amp} \big) + 
Z_S^{(2)} 2 i (p_\alpha p_\nu- p_\alpha q_\nu - p^2 \delta_{\alpha \nu} + p \cdot q \delta_{\alpha \nu}) + \nonumber \\
&& Z_T^{(2)} i (p_\alpha p_\nu - p_\alpha q_\nu - p^2 \delta_{\alpha \nu} + p \cdot q \delta_{\alpha \nu}) + 
\frac{1}{4} \mbox{tr} \big(\gamma_\alpha \big(\frac{\delta O}{\delta \bar \xi(x)}|_{\xi = 0} \big)_{amp}^{(2)} \big) + \nonumber \\
&& Z_{CT}^{(2)} 2 i (p_\alpha q_\nu - p \cdot q \delta_{\alpha \nu} + p^2 \delta_{\alpha \nu} ) - 
Z_{GF}^{(2)} 2 i p_\alpha p_\nu - 
\frac{1}{4} \mbox{tr} \big( \gamma_\alpha \big(\frac{\delta S_{GF}}{\delta \bar \xi(x)}|_{\xi = 0} \big)_{amp}^{(2)} \big) 
\nonumber \\ 
&& + \frac{1}{4} Z_{B_j}^{(2)} \mbox{tr} \big< \gamma_\alpha O B_j \big>^{(0)} = 0 
\label{gammamu}
\eneqa
and 
\beeqa
&& \frac{1}{4} \mbox{tr} \big( \gamma_\alpha \gamma_5 \big( \nabla_\mu S_\mu \big)_{amp}^{(2)} \big) + 
Z_S^{(2)} 2 i p_\rho q_\sigma \varepsilon_{\nu \alpha \rho \sigma} - 
Z_{CT}^{(2)} 2 i p_\rho q_\sigma \varepsilon_{\nu \alpha \rho \sigma} + \nonumber \\
&& \frac{1}{4} \mbox{tr} \big( \gamma_\alpha \gamma_5 \big(\frac{ \delta O}{\delta \bar \xi(x)}|_{\xi = 0} \big)_{amp}^{(2)} \big) 
- \frac{1}{4} \mbox{tr} \big( \gamma_\alpha \gamma_5 \big( \frac{\delta S_{GF}}{\delta \bar \xi(x)}|_{\xi = 0} \big)_{amp}^{(2)} \big) 
+ \nonumber \\
&& \frac{1}{4} Z_{B_j}^{(2)} \mbox{tr} \big< \gamma_\alpha \gamma_5 O B_j \big>^{(0)} = 0 \, .
\label{gammamu5}
\eneqa
More explicitly, the non-trivial part is the computation of the one-loop correction of the projections over 
$\Gamma_r = \left\{ \gamma_\alpha ,\gamma_\alpha \gamma_5 \right\}$ (off-shell regime), for
$ \mbox{tr} \big( \Gamma_r \big(\nabla_\mu S_\mu \big)^{(2)}_{amp} \big)$ (12 diagrams), 
$  \mbox{tr} \big( \Gamma_r \big(\frac{\delta S_{GF}}{\delta \bar \xi(x)}|_{\xi = 0} \big)_{amp}^{(2)} 
\big) $ (12 diagrams) and 
$ \mbox{tr} \big(\Gamma_r \big(\frac{\delta O}{\delta \bar \xi(x)}|_{\xi = 0} \big)_{amp}^{(2)} \big)$ 
(4 diagrams), for a total of 28 Feynman diagrams for the determination of $Z_T^{(2)}$ 
(see Ref.~\cite{feo2} for details).

The determination of $Z_T^{(2)}$ require in principle the knowledge of which operators in 
$ Z^{(2)}_{B_j} \mbox{tr} \big< \Gamma_r O B_j \big>^{(0)} $ should be included. 
One can demonstrate \cite{feo2} that the calculation of $Z_T^{(2)}$ does not involve any 
of this operators. Another point here is that, even
in the continuum limit, Lorentz breaking terms appears in Eqs.~(\ref{gammamu},\ref{gammamu5}),
but imposing the on-shell condition on the gluino mass, (once the value of $Z_T^{(2)}$ has been
determined) they dissapear and the continuum WTi is recovered.
The final value is, $Z_T^{(2)}|_{1-loop} = 0.332$, which  gives good agreement with the one 
in \cite{farchioni}.

\section{OUTLOOK}
A big effort has been made in order to describe supersymmetry on the lattice.
Wilson fermions have been used in realistic computations with nice results. 
Improved chiral fermions results are starting too. 
New interesting proposals, \cite{kaplan}, are waiting for numerical applications.

\vskip1cm
\noindent

\section*{ACKNOWLEDGEMENTS}

I wish to thank the organisers of the LC03, in particular, S.~S.~Pinsky, S. J. Brodsky, S.~Dalley, J.~R.~Hiller, 
M.~Karliner, W.~J.~Stirling, for having invited me to such a pleasant and stimulating workshop.
It is also a pleasure to thank M.~Bonini, M.~Beccaria, M.~Campostrini, R.~De~Pietri, H.~J.~Pirner,
F.~Sannino, for interesting discussions.

\end{document}